\documentclass[showpacs,preprintnumbers,amsmath,amssymb,prl,superscriptaddress,twocolumn,longbibliography,footinbib]{revtex4-1}

\usepackage{natbib}
\usepackage{txfonts}
\usepackage{graphicx}% Include figure files
\usepackage{dcolumn}% Align table columns on decimal point
\usepackage{bm}% bold math
\usepackage{hyperref}
\usepackage{color}
\usepackage{subfigure}  % use for side-by-side figures
\def\comment#1{}

\DeclareMathAlphabet\mathbfcal{OMS}{cmsy}{b}{n}

\def\slashchar#1{\setbox0=\hbox{$#1$}           % set a box for #1
	\dimen0=\wd0                                 % and get its size
	\setbox1=\hbox{/} \dimen1=\wd1               % get size of /
	\ifdim\dimen0>\dimen1                        % #1 is bigger
	\rlap{\hbox to \dimen0{\hfil/\hfil}}      % so center / in box
	#1                                        % and print #1
	\else                                        % / is bigger
	\rlap{\hbox to \dimen1{\hfil$#1$\hfil}}   % so center #1
	/                                         % and print /
	\fi}                                         %

\def\nablab{{\mbox{\boldmath $\nabla$}}}
\def\varphib{{\mbox{\boldmath $\varphi$}}}
\def\gammab{{\mbox{\boldmath $\gamma$}}}

\begin{document}

%\title{Axion-induced Electrically Polarized Vortices in Superconductor-Strong Topological Insulator Systems}
%
%\title{Dyon-like vortex and unconventional quantization of angular momentum at superconductor - topological insulator interfaces}
%
%\title{\jb {Continuously tunable angular momenta quanta at topological insulator interfaces} }
%
\title{Fractional angular momentum at topological insulator interfaces} 
\author{Flavio S. Nogueira}
%\email{flavio.nogueira@fu-berlin.de}
\affiliation{Institute for Theoretical Solid State Physics, IFW Dresden, Helmholtzstr. 20, 01069 Dresden, Germany}
\affiliation{Institut f{\"u}r Theoretische Physik III, Ruhr-Universit\"at Bochum,
Universit\"atsstra\ss e 150, 44801 Bochum, Germany}

\author{Zohar Nussinov}
%\email{zohar@wuphys.wustl.edu}
\affiliation{Physics Department, CB 1105, 
Washington University, 1 Brookings Drive, St. Louis, MO 63130-4899}

\author{Jeroen van den Brink}
\affiliation{Institute for Theoretical Solid State Physics, IFW Dresden, Helmholtzstr. 20, 01069 Dresden, Germany}
%\affiliation{Department of Physics, Harvard University, Cambridge, Massachusetts 02138, USA}
\affiliation{Institute for Theoretical Physics, TU Dresden, 01069 Dresden, Germany}

\date{Received \today}

%Nature Physics: The maximum title length is 15 words. The abstract is typically 150 words and is unreferenced;

\begin{abstract}
Recently two fundamental topological properties of a magnetic vortex at the interface of a superconductor (SC) and a strong topological insulator (TI) have been established: the vortex carries both a Majorana zero-mode relevant for topological quantum computation and, for a time-reversal invariant TI, a charge of $e/4$. 
This fractional charge is caused by the axion term in the electromagnetic Lagrangian of the TI. 
Here we determine the angular momentum $J$ of the vortices, which in turn determines their mutual statistics.
Solving the axion-London electrodynamic equations including screening in both SC and TI, we find that the elementary quantum of angular momentum of the vortex is $-n^2\hbar/8$, where $n$ is the flux quantum of the vortex line. 
Exchanging two elementary fluxes thus changes the phase of the wavefunction by $-\pi/4$.
\end{abstract}

%\pacs{64.70.Tg, 11.10.Kk, 11.15.Ha}
\maketitle

In our three-dimensional world, a particle must either be a boson or a fermion. Exchanging two identical particles either leaves the wavefunction invariant (as it does for bosons) or induces a sign-change (for fermions). 
By the spin-statistics theorem, these two distinct possibilities are related to the intrinsic angular momentum of bosons (or fermions) being an integer (or
a half odd integer) multiple of the fundamental quantum unit of $\hbar$.
When restricted to two spatial dimensions (2D), quantum statistics become much richer. In principle, particles can exist in 2D that, insofar as quantum statistics are concerned, lie ``in between" bosons and fermions.
Such particles, dubbed ``anyons'', are predicted to have very peculiar physical properties \cite{Leinaas1977,Wilczek-anyon}. 
Exchanging two identical anyons may produce a change of the phase of the wavefunction that is anywhere between zero and $\pi$; the intrinsic angular momentum number of anyons need not be an integer or half-odd multiple of $\hbar$. 
Topological quantum computation relies on the ability to create and manipulate anyons with nontrivial (particularly, more complex non-Abelian) statistics \cite{Pachos-book,Nayak-RMP}.  

Theoretically, anyons can emerge as composite particles formed by a bound state between a 2D fermion and a flux tube which provides an additional 
Aharonov-Bohm (AB) phase to exchanged particles~\cite{Wilczek-anyon}.
However, in conventional electromagnetism governed by the Maxwell field equations, the canonical angular momentum {\it cannot} be fractional \cite{Jackiw-Redlich,Forte1992}.
Anyons play a prominent role in effective theories of the fractional quantum Hall effect \cite{Nayak-RMP} wherein fermions are coupled to {\it auxiliary} emergent gauge-fields. 
In these phenomenological field theories, the fractional angular momentum arises from a topological (Chern-Simons) term in the effective Lagrangian and field equations. 

The last decade has seen an explosion in experimental discoveries of new classes of materials in which topological effects come to life \cite{Hasan2010}.  A strong topological insulator (TI) is characterized by a topologically protected surface state that consists of a single Dirac cone. 
Electromagnetic fields that enter or exit the TI couple to these Dirac fermions, which, interestingly, gives rise to an additional, non-perturbative, topological term in the Maxwell Lagrangian. This (so-called axion) term couples the electric ($\bf E$) and magnetic ($\bf B$) fields and is given by ${\cal L}_{\rm axion}=\alpha\theta/(4\pi^2){\bf E}\cdot{\bf B}$ \cite{Witten,Wilczek}. 
Here, $\alpha$ is the fine-structure constant and $\theta$ the coupling constant. For a time-reversal invariant strong TI, $\theta=\pi$. Hallmarks of the axion term have recently been detected in Bi$_2$Se$_3$, a prominent TI material\cite{Wu2016}.

\begin{figure}
	\centering
	\begin{tabular}{c}
		\hspace{0.5cm}(a)\includegraphics[width=0.45\columnwidth]{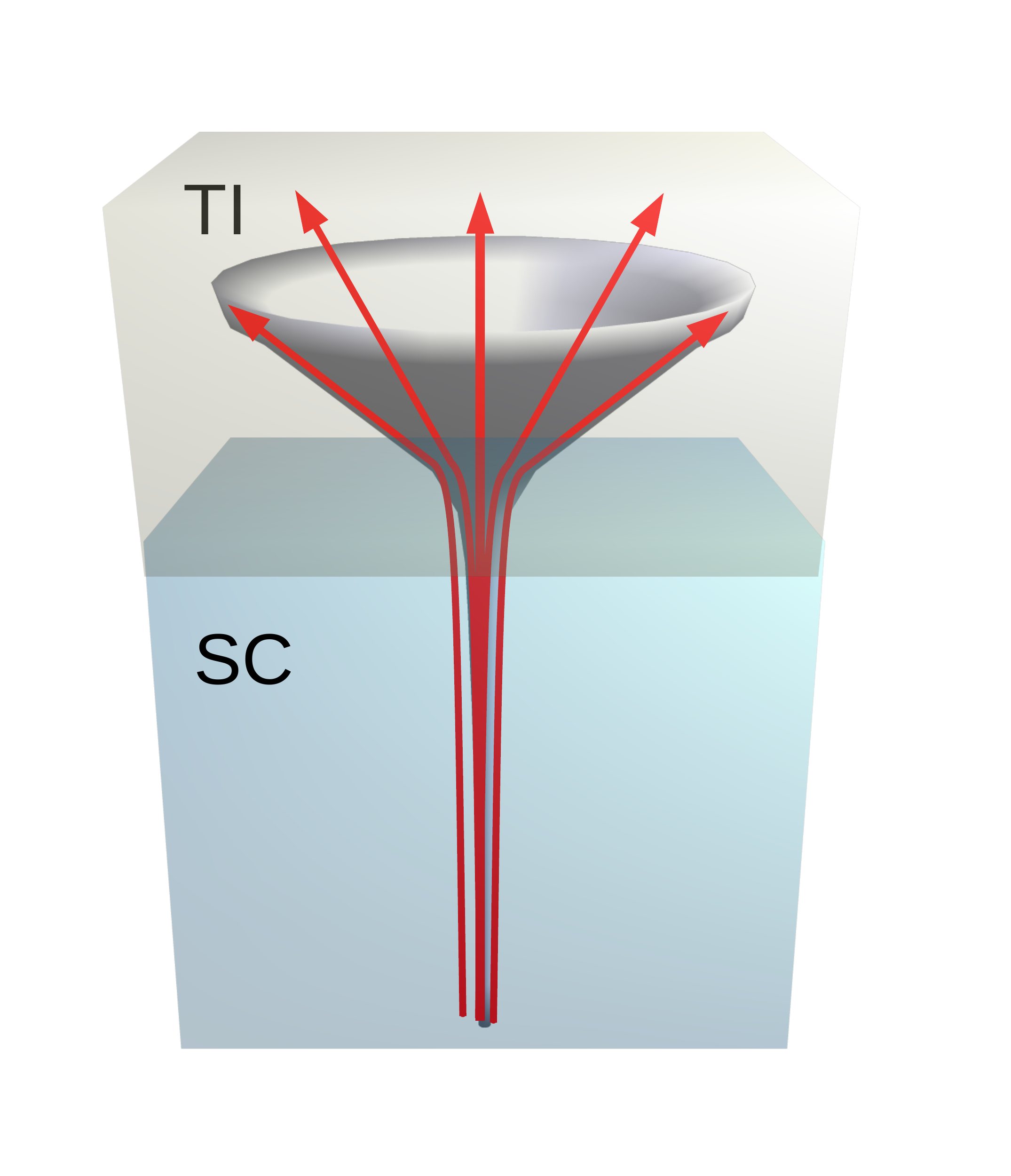}
		\hspace{-0.6cm}(b)\includegraphics[width=0.48\columnwidth]{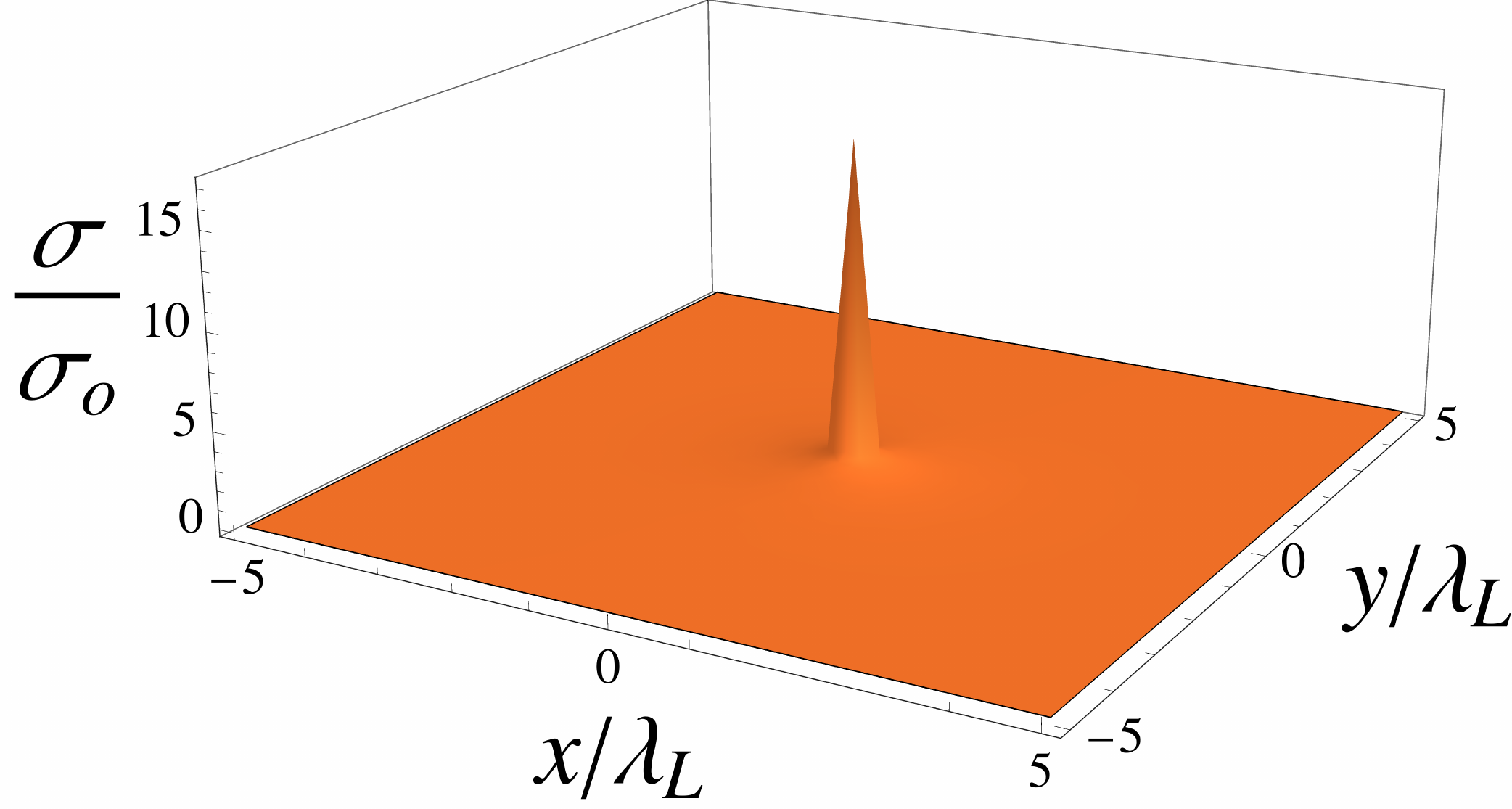} \\[6pt]
		(c)\hspace{-0.5cm}\includegraphics[width=0.45\columnwidth]{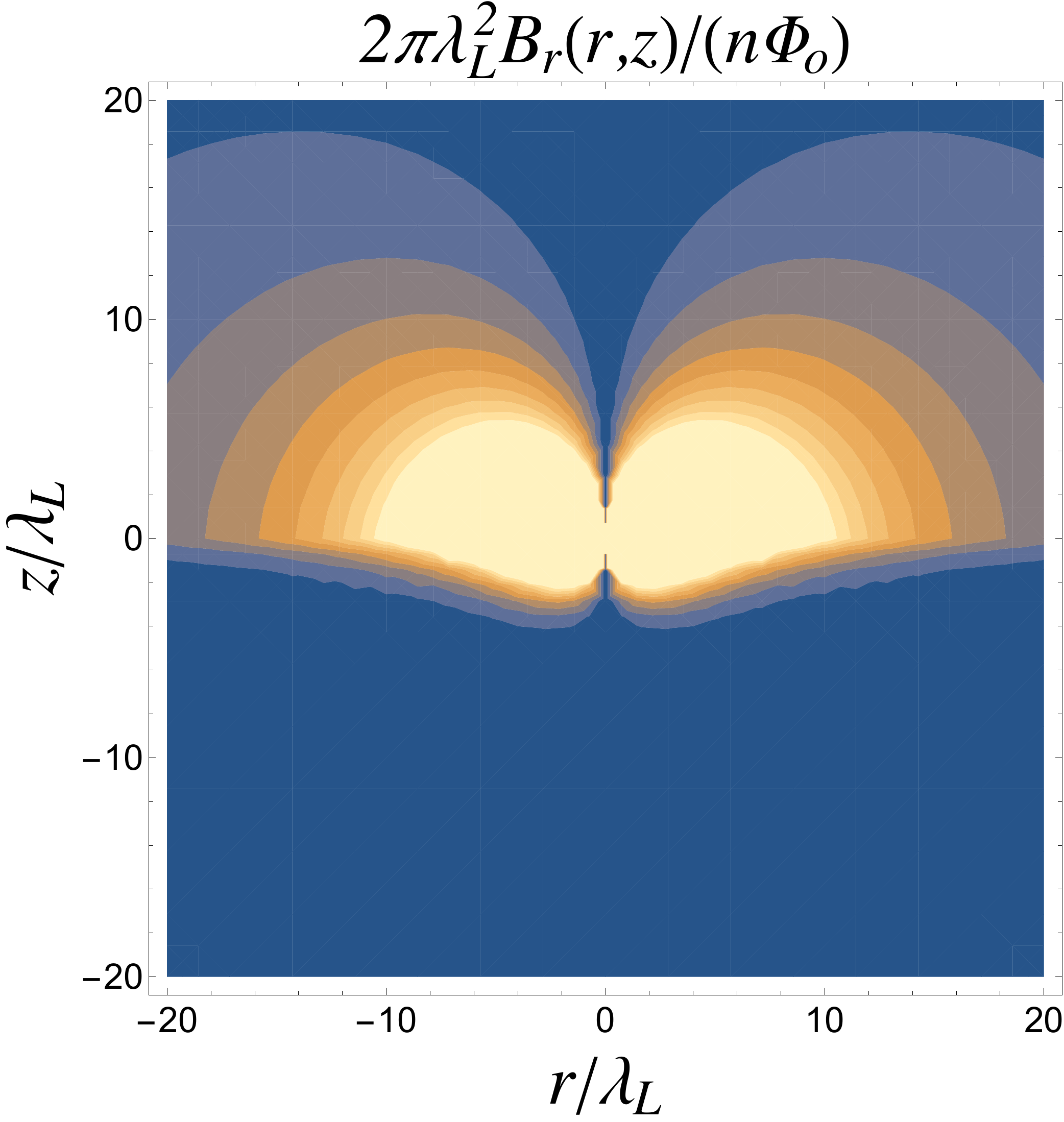} ~~~~%\\[6pt]
		(d)\hspace{-0.5cm}\includegraphics[width=0.45\columnwidth]{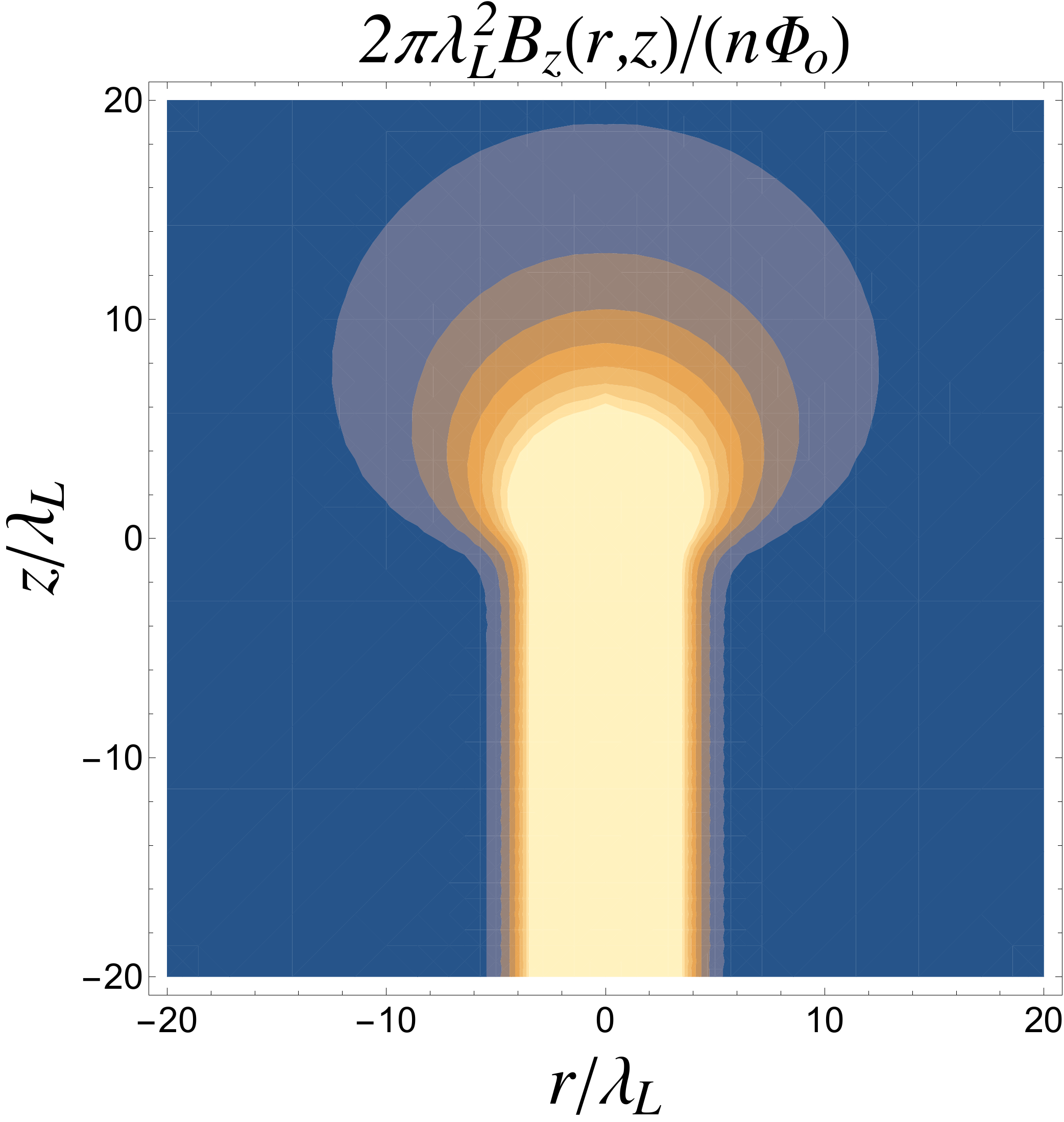} \\[6pt]
		(e)\hspace{-0.5cm}\includegraphics[width=0.45\columnwidth]{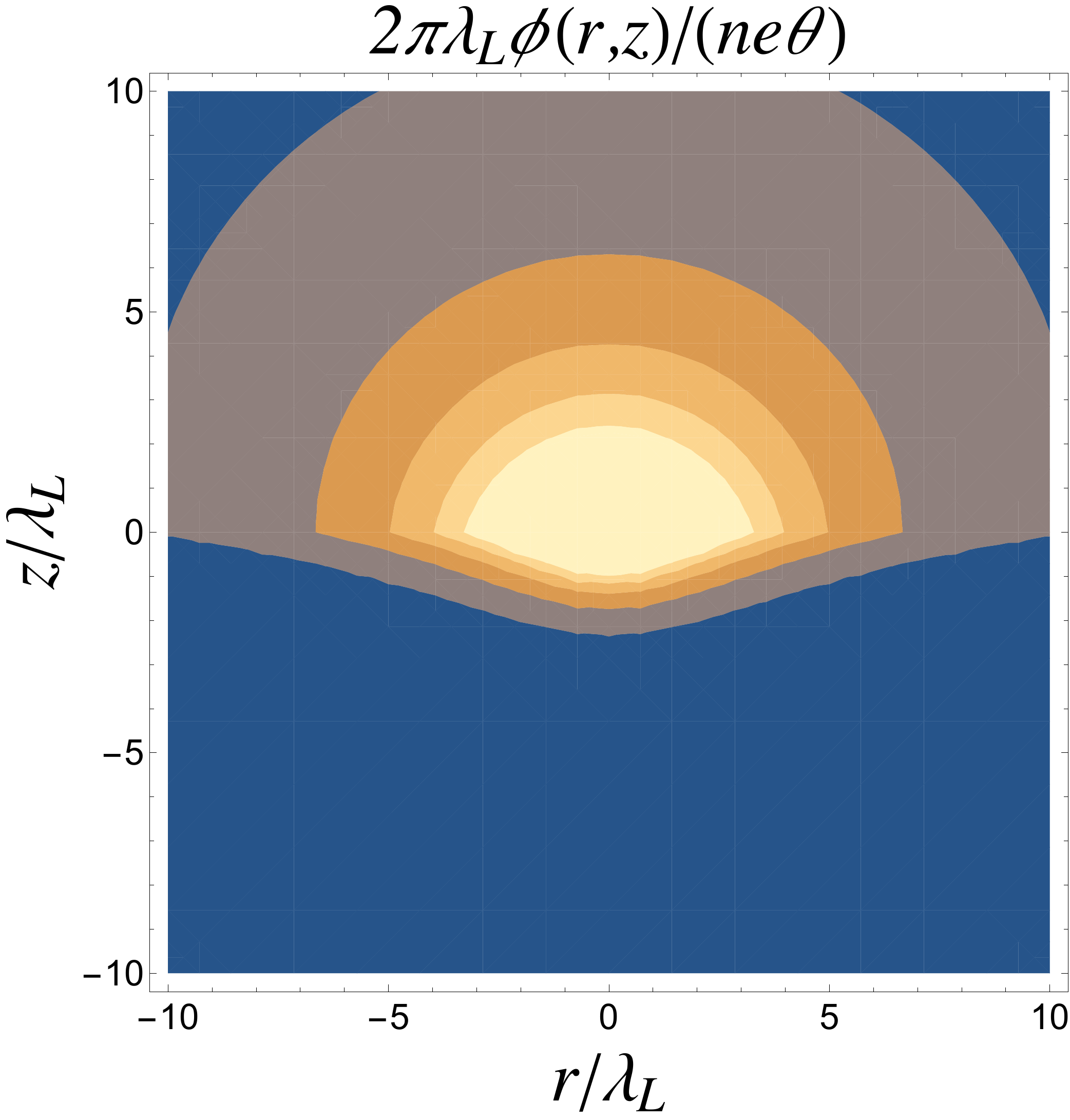}
		%(e)\hspace{-0.5cm}\includegraphics[width=0.45\columnwidth]{Er} ~~~~%\\[6pt]
		%(f)\hspace{-0.5cm}\includegraphics[width=0.45\columnwidth]{Ez}
	\end{tabular}
	\caption{(a) Schematic representation of a magnetic vortex in a topological insulator (TI) - superconductor (SC) heterostructure. At the interface, where the vortex line ends, stray fields similar the field of a magnetic monopole appear which however do not violate the Maxwell equation $\nablab\cdot{\bf B}=0$ (see main text). Panel (b) shows the axion-induced 
		electric charge density $\sigma$ divided by $\sigma_0=ne\theta/(8\pi^2\lambda_L^2)$, summing up to a total charge $e/4$ for 
		$\theta=\pi$; (c) and (d): contour plots of the radial and $z-$components of magnetic field associated to the vortex;
		(e) contour plot of electric potential induced by the magnetic vortex. Due to the axion term, the magnetic vortex line becomes a source of electric field. 
		}
		%(e) and (f): contour plots of the induced radial and $z$-components of the electric field. Due to the axion term, the magnetic vortex line becomes a source of electric field.}
	\label{Fig:Magnetic-field}
\end{figure}

It was recently demonstrated that the axion term binds an {\it electric} charge of  $e/4$ to a {\it magnetic} vortex (see Fig. \ref{Fig:Magnetic-field}-a) at the interface of a conventional type II superconductor (SC) and a time-reversal invariant TI \cite{Nogueira-Nussinov-van_den_Brink-PRL,Nogueira-Nussinov-van_den_Brink-PRD}.
At the interface, the magnetic flux (i.e., the magnetic vortex emerging from the SC) becomes endowed with an electric charge via an elementary realization of the Witten effect \cite{Witten}.
The vortex has also been shown to carry a Majorana zero-mode due to the SC proximity effect in the TI \cite{Fu-Kane-2008,Vishwanath-2011}.
A principal result of the current work is the demonstration that such vortices in addition carry a canonical angular momentum $J$ that is not an integer multiple of $\hbar/2$. This rather unconventional feature is triggered by the topological axion term in the  field equations governing the electromagnetic response of the TI. In this way the results of the present work can be viewed as providing a realization of a bosonic 
TI \cite{Senthil-Vishwanath_PhysRevX.3.011016}. In addition, fractional statistics can be realized by means of a gauge field corresponding 
to the actual electromagnetic field. A statistical Witten effect considered earlier in the context of bosonic TIs required a compact 
gauge field \cite{Metlitski-Kane-Fisher_PhysRevB.88.035131}, which naturally leads to magnetic monopole excitations. 
However, even in this context, duality arguments show an equivalence to a system featuring electromagnetic vortices 
rather than monopoles \cite{Metlitski-Vishwanath_PhysRevB.93.245151,Nogueira-Nussinov-van_den_Brink-PRD}.   

In what follows, we derive the detailed electromagnetic field profile of the vortex by explicitly solving the axion-London electrodynamic equations including screening in both the SC and  the TI.
In particular, the electromagnetic field induced by the vortex depends on the screening properties of both  the SC and the TI via the London penetration depth and dielectric constant $\epsilon$. 
This subtle behavior arises because the (effectively 2D) interface between the TI and SC that we consider is embedded in a 3D polarizable medium.
We establish that the total angular momentum is given by $J_z^{\rm tot}=-n^2\hbar\theta/(8\pi)$, where $n$ is the quantum of the 
vortex flux. 
Interestingly, the fractionalization of $J_z^{\rm tot}$ is solely a property of the electromagnetic fields matching at the interface. Our results do not hinge on the existence of a SC proximity effect on the surface of the TI. 

% \begin{figure}
% 	\begin{tabular}{c}
%		\includegraphics[width=0.5\columnwidth]{fluxtube.pdf}
%	\end{tabular}
% 	\caption{Schematic representation of a vortex at the interface of a superconductor (SC) and a topological insulator (TI)}
 %	\label{Fig:Electric-pot-cd}
 %\end{figure}

{\it London-axion electrodynamics ---}  
We consider a SC-TI interface at $z=0$ in the London regime (see Fig.~\ref{Fig:Magnetic-field}) and  employ cylindrical coordinates ${\bf R}=({\bf r},z)$. The SC occupies the lower half of space, $z<0$, with the  TI lying in the region above it. Thus, $\theta(z)=\theta={\rm const}$ for $z\geq 0$, vanishing otherwise. 
In the static limit, Maxwell's equations read
\begin{eqnarray}
\label{Eq:Gauss-SC}
\nablab\cdot\left[\epsilon(z){\bf E}({\bf r},z)-\frac{\alpha\theta(z)}{\pi}{\bf B}({\bf r},z)\right]=4\pi\rho({\bf r},z), \\
\label{Eq:Ampere-SC}
\nablab\times\left[{\bf B}({\bf r},z)+\frac{\alpha\theta(z)}{\pi}{\bf E}({\bf r},z)\right]=\frac{4\pi}{c}{\bf j}({\bf r},z),
\end{eqnarray} 
 along with the equations $\nablab\cdot{\bf B}=0$ and $\nablab\times{\bf E}=0$. 
Here, the dielectric constant
$\epsilon(z)=\epsilon={\rm const}$ in the region $z>0$ inside the TI, while for $z<0$, inside the SC, $\epsilon(z)=1$. The charge density and current are given by
\begin{equation}
\label{Eq:cd}
\rho({\bf r},z)=-\frac{(2e)^2\rho_s}{mc^2}\phi({\bf r},z),
\end{equation}
\begin{equation}
\label{Eq:currd}
{\bf j}({\bf r},z)=2e\rho_s{\bf v}_s({\bf r},z).
\end{equation}
$\rho_s$ is the superfluid density, $m$ is the mass of the Cooper pair and $\phi$ is the electric potential, and
\begin{equation}
\label{Eq:vs}
{\bf v}_s({\bf r},z)=\frac{1}{m}\left[\hbar\nablab\varphi_T-\frac{2e}{c}{\bf A}({\bf r},z)\right],
\end{equation}
is the superfluid velocity of the SC.
For $z>0$, inside the TI, $\nablab\times{\bf B}=0$ and $\nablab\cdot{\bf E}=0$.   

In Eq. (\ref{Eq:vs}) the longitudinal phase gradient is gauged away while the transverse phase gradient is given by 
\begin{equation}
\label{Eq:grad-T}
\nablab\varphi_T=\frac{1}{2}\int_{L}\frac{d\gammab\times({\bf R}-\gammab)}{|{\bf R}-\gammab|^3},
\end{equation}
where the line integral is along the vortex line. For simplicity, we consider a single straight vortex line centered about ${\bf{r} }=0$ lying parallel to the $z$-axis.  
In the absence of the TI, the SC would occupy all space, and the vortex line would extend over $z'\in(-\infty,\infty)$. In this case, a simple calculation using $\gammab=z'\hat{\bf z}$ in 
Eq. (\ref{Eq:grad-T}) yields $\nablab\varphi_T=(n/r)\hat{\varphib}$, $n\in\mathbb{Z}$.  
As a result, the standard London equation holds and its solution yields \cite{A-vortices,NO-vortices}, 
\begin{equation}
\label{Eq:London-sol}
{\bf B}({\bf r})=\frac{n\Phi_0}{2\pi}m_L^2K_0(m_Lr)\hat {\bf z}.
\end{equation}
Here, $\Phi_0= hc/(2e)$ is the fundamental flux of the vortex in the superconductor, $K_0(u)$ is a modified Bessel function of second kind, and
$m_L^2=16\pi e^2\rho_s/(mc^2)=1/\lambda_L^2$, with $\lambda_L$ being the penetration depth.

In our case of an SC-TI structure, the integral in Eq. (\ref{Eq:grad-T}) has to be performed over  $z'\in(-\infty,0)$. We obtain, 
\begin{equation}
\label{Eq:nablavT}
\nablab\varphi_T=\frac{n}{2r}\left(1-\frac{z}{\sqrt{r^2+z^2}}\right)\hat{\varphib},
\end{equation}
which in spherical coordinates reads, $\nablab\varphi_T=n\hat{\varphib}(1-\cos\vartheta)/(R\sin\vartheta)$. We recognize this as 
the gauge potential for a magnetic monopole of charge $n/2$ with a Dirac string extending over the negative $z$-axis.   
 Since ${\bf A}(r,z)=A(r,z)\hat{\varphib}$, the monopole contribution is removed by  
 performing the gauge transformation $A\to A-n\Phi_0(1+z/\sqrt{r^2+z^2})/(4\pi r)$, such that 
 Eq. (\ref{Eq:vs}) becomes,
 \begin{equation}
 \label{Eq:vs-1}
 {\bf v}_s=\frac{1}{m}\left(\frac{n\hbar}{r}-\frac{2e}{c}A\right)\hat \varphib.
 \end{equation}
 This is similar to the case of an infinitely long vortex line, except that the vector potential ${\bf A}(r,z)=A(r,z) \hat{\varphib}$ depends on $z$ to account for the presence of 
 the interface at $z=0$.

 {\it Explicit solution for a single vortex ---} 
  In cylindrical coordinates, Eq. (\ref{Eq:Ampere-SC}) becomes the
 partial differential equation
 \begin{equation}
 \label{Eq:PDE}
 -\frac{\partial^2A}{\partial r^2}-\frac{1}{r}\frac{\partial A}{\partial r}+\frac{A}{r^2}-\frac{\partial^2A}{\partial z^2}
 +m_L^2A=\frac{m_L^2n\Phi_0}{2\pi r}.
 \end{equation}
 This equation has to be solved by imposing continuity of $A(r,z)$ at $z=0$ together with the boundary condition, 
 \begin{equation}
 \label{Eq:BC1}
 \left.\frac{\partial A}{\partial z}\right|_{z=+\eta}-\left.\frac{\partial A}{\partial z}\right|_{z=-\eta}=\frac{\alpha\theta}{\pi}\left.
 \frac{\partial\phi}{\partial r}\right|_{z=0}, 
 \end{equation}
 where $\eta\to +0$.    
 An additional boundary condition requires that the magnetic field approach Eq. (\ref{Eq:London-sol}) as $z\to -\infty$. It is easy to show that the latter condition corresponds to 
 $A(r,z=-\infty)=n\Phi_0[1/r-m_LK_1(m_Lr)]/(2\pi)$. 
 
 With these boundary conditions, the solution to (\ref{Eq:PDE}) is 
 \begin{equation}
 \label{Eq:A}
 A(r,z)=\frac{n\Phi_0m_L^2}{2\pi}\int_0^\infty dp\frac{J_1(pr)a(p,z)}{p^2+m_L^2}.
 \end{equation}
 Here, $J_1(x)$ is a Bessel function of first kind. The function
 $a(p,z)=1+e^{z\sqrt{p^2+m_L^2}}[f(p)-1]$ for $z\leq 0$; otherwise $a(p,z)=f(p)e^{-pz}$. The function $f(p)$ (see SI) has to be determined 
 with the help of Eq. (\ref{Eq:Gauss-SC}) along with the boundary condition (\ref{Eq:BC1}). The above solution fulfills $\nablab\times{\bf B}=0$ 
 for $z>0$, as it should.    
 Note that 
 $\nablab\cdot{\bf B}=0$ holds, since the vortex line plays the role of a (screened) Dirac string.   
 Indeed, we note the large distance ($m_Lr\gg 1$) behavior, 
 \begin{equation}
 \label{Eq:A-largedist}
 A(r,z)\approx
 \begin{cases}
 \frac{n\Phi_0}{2\pi r}, & z \leq 0\\
 \frac{n\Phi_0}{2\pi r}\left(1-\frac{z}{\sqrt{r^2+z^2}}\right)  & z>0.
 \end{cases}
 \end{equation}
 This is the gauge field of a thin solenoid of quantized flux ($n \Phi_0$) for $z\leq 0$ and the gauge field of a monopole for $z>0$.
 
  Equation (\ref{Eq:Gauss-SC}) can now be solved  using the  
  boundary conditions
  $\phi({\bf r},z=+\eta)=\phi({\bf r},z=-\eta)$ and  
  %\end{equation} 
  %
  \begin{equation}
  \label{Eq:BC2}
  \left.\frac{\partial\phi}{\partial z}\right|_{z=-\eta}-\epsilon\left.\frac{\partial\phi}{\partial z}\right|_{z=+\eta}=
  \frac{\alpha\theta}{\pi r}\left.\frac{\partial (rA)}{\partial r}\right|_{z=0}=\frac{\alpha\theta}{\pi}B_z(r,z=0). 
  \end{equation}
  The above boundary condition reflects the fact that Eq. (\ref{Eq:Gauss-SC}) implies that 
  $\epsilon(z)E_z({\bf r},z)-(\alpha\theta(z)/\pi)B_z({\bf r},z)$ is continuous at $z=0$, which in turn leads to a 
  discontinuity in $E_z$ at $z=0$. 
  This is reminiscent of the 
  continuity at $z=0$ (assuming the same geometry) of the normal component of the 
  electric displacement field ${\bf D}({\bf r},z)=\epsilon(z){\bf E}({\bf r},z)$ in 
  absence of the so called "free charges". In our case it is the vector $\epsilon(z){\bf E}({\bf r},z)-(\alpha\theta(z)/\pi){\bf B}({\bf r},z)$ that 
  plays the role of ${\bf D}$. 
  
  It is now straightforward to obtain the solution, 
  \begin{equation}
  \label{Eq:phi}
  \phi(r,z)=\frac{ne\theta m_L^2}{2\pi}\int_0^\infty dp\frac{pJ_0(pr)F(p,z)}{\left(\sqrt{p^2+m_L^2}+\epsilon p\right)(p^2+m_L^2)},
  \end{equation}
  where $F(p,z)=f(p)e^{z\sqrt{p^2+m_L^2}}$ for $z\leq 0$ and $F(p,z)=f(p)e^{-pz}$ otherwise. We can verify that Eq. (\ref{Eq:phi}) yields 
  $\nablab\cdot{\bf E}=0$ for $z>0$. 
  
The boundary condition (\ref{Eq:BC1}) enables us to determine $f(p)$. Its explicit form displays a very weak dependence on 
$\theta$, departing from the $\theta$-independent expression only by a correction $\sim(\alpha\theta/\pi)^2\sim 10^{-5}$. 
Thus, for all practical purposes the $\theta$-dependence of $f(p)$ can be ignored, yielding  
$f(p)=\sqrt{p^2+m_L^2}/\left(p+\sqrt{p^2+m_L^2}\right)$. It is interesting to note that this function is independent of $\epsilon$ (for details, see SI).  
%
%\begin{equation}
%\label{Eq:f}
%f(p)=\frac{p\left(p+\epsilon \sqrt{p^2+m_L^2}\right)+m_L^2}{p\left\{[1+\epsilon+(\alpha\theta/\pi)^2]p+(1+\epsilon)\sqrt{p^2+m_L^2}\right\} +m_L^2}.
%\end{equation}
Note that $a(0,z)=1$, so that the usual flux quantization holds,  
%\begin{equation}
$
\int d^2r B_z(r,z)=n\Phi_0a(0,z)=n\Phi_0.
$
%\end{equation} 
%Moreover, for all practical purposes we will neglect the term $(\alpha\theta/\pi)^2$ in the denominator of (\ref{Eq:f}), 
%since it is $\sim 5\times 10^{-5}$ for $\theta=\pi$, being therefore negligible compared to $1+\epsilon>2$. 

{\it Charge bound to vortex ---} 
Equation (\ref{Eq:Gauss-SC}) with the  above fields implies that the vortex line carries a quantized fractional charge. Indeed, 
noting that for any function ${\cal F}$ of $p=|{\bf p}|=\sqrt{p_x^2+p_y^2}$ and $z$, 
\begin{equation}
\int_0^\infty\frac{dp}{2\pi}pJ_0(pr){\cal F}(p,z)=\int\frac{d^2p}{(2\pi)^2}e^{i{\bf p}\cdot{\bf r}}{\cal F}(p,z),
\end{equation}
we easily obtain that induced fractional charge is independent of $\epsilon$ and is given by, 
\begin{equation}
\label{Eq:Q}
Q=\int_{-\infty}^0 dz\int d^2r\rho(r,z)=-\frac{ne\theta}{4\pi},
\end{equation}
behaving in this way similarly to dyons in the Witten effect \cite{Witten}.  
%Note that from the boundary condition (\ref{Eq:BC2}) that as a consequence of the axion term,  there is also a surface "free charge" induced by the normal component of the magnetic field. When integrate over the entire interface ($z=0$) plane, it yields precisely $-Q$. Thus,  the {\it total} charge (i.e., the sum of the bulk and surface charge contributions) vanishes \cite{Nogueira-Nussinov-van_den_Brink-PRL,Nogueira-Nussinov-van_den_Brink-PRD} such that charge conservation is not violated.  
 Thus, the axion term causes the magnetic vortex line to become electrically polarized. 
 
In Fig. \ref{Fig:Magnetic-field}-(b)  we show the charge density of the vortex at the interface. Here and in all other panels of Fig. \ref{Fig:Magnetic-field}, we have used $\epsilon=20$, corresponding to the bulk dielectric constant of HgTe~\cite{Capper1994}. The  contour plots of the vortex magnetic field components are shown in panels (c) and (d) of Fig. \ref{Fig:Magnetic-field}. The stray fields in the $z>0$ region ensure that the magnetic monopole contribution cancels out enforcing in this way the Maxwell equation $\nablab\cdot{\bf B}=0$ \cite{Brandt-1981}.  
The induced charge density shows that the bound $e/4$ charge lying at the SC-TI interface is localized, within a radius of $\lambda_L$, to the vortex core.

Fig. \ref{Fig:Magnetic-field}-(e) shows the {\it electric} potential that is induced by the magnetic vortex at the SC-TI interface and screened by the rest of the system.  It clearly demonstrates that due to the axion term the vortex line flux becomes the source of the electric field. In the SI \cite{SI}, the axion-induced electric field components are shown - the radial electric field profile is similar to the radial magnetic field profile in Fig. \ref{Fig:Magnetic-field}-(b). The field profile of the $z-$component of the electric field exhibits a discontinuous behavior at $z=0$, which is a consequence of the discontinuity of the normal component of the electric field across the interface, as expressed mathematically in Eq. (\ref{Eq:BC2}). 

{\it Angular momentum ---}  
We now turn to our central result concerning the appearance of fractionalized angular momenta of the dyons.
As is well known \cite{Schwinger-1969}, 
dyons, which are dipoles constituted by electric and magnetic poles, have an angular momentum proportional to the 
electric charge times the flux from the magnetic monopole. Julia and Zee \cite{Julia-Zee} have shown that an 
Abrikosov-Nielsen-Olesen vortex \cite{A-vortices,NO-vortices} would also have an angular momentum of a similar form, provided 
the vortex line has also an electric potential associated to it. However, it turns out that such a vortex solution would imply an infinite 
vortex energy per unit length, since the electric potential would behave logarithmically at short distances, so there must be no charge and 
the total angular momentum vanishes. The axion term offers a way out of this problem, since it gives a fractional charge to the vortex and 
the electric contribution to its energy is finite, as can easily be verified using Eq. (\ref{Eq:phi}).  We will now calculate exactly the angular momentum of 
the vortex and shows that its fractional charge implies a unconventional quantization of the angular momentum. 

The total angular momentum is given by $\mathbfcal{J}={\bf L}+{\bf J}$, where 
\begin{equation}
\label{Eq:m-angmom}
{\bf L}=\int d^3R({\bf R}\times\mu{\bf v}_s),
\end{equation}
is the mechanical angular momentum and ${\bf J}$  
%Having explicit and exact expressions for electric and magnetic fields related to the vortex we now can determine the intrinsic angular momentum distribution of the vortex structure. Due to the axion term, the electromagnetic fields give rise to a {\it topological} contribution to the total angular momentum that is absent is standard Maxwellian electromagnetism~\cite{Jackiw-Redlich,Forte1992}.
%
%The vortex can in principle also carry a mechanical (or orbital) angular momentum, but it is straightforward to show that it does not depend on $\theta$ in leading order $\alpha$ with $\sim(\alpha\theta/\pi)^2\sim 10^{-5}$ corrections beyond that, see \cite{SI}, so that it is irrelevant for the topological angular momentum.
%
%The total electromagnetic field contribution ${\bf J}$ to the angular momentum can be computed by using the standard formula 
%
%\begin{equation}
%\label{Eq:em-angmom}
%{\bf J}=\frac{1}{4\pi c}\int d^3R[{\bf R}\times({\bf E}\times{\bf B})].
%\end{equation}
is the angular momentum of the electromagnetic field, which for static fields can be written as, 
\begin{equation}
\label{Eq:em-angmom-1}
{\bf J}=\frac{1}{4\pi c}\int d^3R(\nablab\cdot{\bf E})({\bf R}\times{\bf A}). 
\end{equation}
In Eq. (\ref{Eq:m-angmom}) $\mu$ is the mass density of 
the charged superfluid, which is given by $\mu({\bf R})=m\rho({\bf R})/(2e)$.  The expression for ${\bf J}^{\rm tot}$ follows more generally 
from the $T_{0i}$ component of the symmetric energy-momentum tensor via the integral of the expression $\epsilon_{ijk}x_jT_{0k}$. 
Since the energy momentum tensor is by definition the conserved tensor current in response to changes in the metric, it does not 
depend explicitly on the axion term, which is a topological term.  

Rotational invariance in the plane perpendicular to the vortex line implies that only the $z$-component of the angular momentum is nonzero. Thus, 
using Eqs. (\ref{Eq:cd}) and (\ref{Eq:vs-1}), we obtain, 
\begin{equation}
L_z=\frac{2e\rho_s}{mc^2}\int_{-\infty}^0 dz\int d^2r\phi(r,z)\left[\frac{2e}{c}rA(r,z)-n\hbar\right], 
\end{equation}
while from Eq. (\ref{Eq:em-angmom-1}) we have, 
\begin{equation}
J_z=-\frac{m_L^2}{4\pi c}\int_{-\infty}^0 dz\int d^2r~r\phi(r,z)A(r,z),
\end{equation}
since $\nablab\cdot{\bf E}=4\pi\rho=-m_L^2\phi$ for $z<0$, vanishing otherwise. Therefore, adding the two contributions above we obtain (recall that 
$m_L^2=16\pi e^2/(mc^2)$),
\begin{eqnarray}
\label{Eq:J-tot}
{\cal J}_z=-\frac{2e\rho_sn\hbar}{mc^2}\int_{-\infty}^0 dz\int d^2r\phi(r,z)
%\nonumber\\
%&=&\frac{n\hbar}{2e}\int_{-\infty}^0 dz\int d^2r\rho(r,z)
=\frac{n\Phi_0Q}{2\pi c}.
\end{eqnarray}
In the absence of the TI, and therefore from the axion term, 
$\phi$ is obtained from the solution of the London equation, $-\nabla^2\phi+m_L^2\phi=0$, which for an infinite system has 
the form, $\phi(r)=-4\lambda K_0(m_Lr)$, where $\lambda$ is the charge per unit length of the vortex.  
This leads to an infinite energy 
per unit length, and therefore only the trivial solution with zero charge can exist \cite{Julia-Zee}, which in turn implies that 
 ${\cal J}_z$ vanishes. 
In contrast, thanks to the axion term, the case discussed above  
features a magnetic vortex flux which becomes the source of an electric field and a finite energy solution is obtained \cite{SI}.  

Inserting Eq. (\ref{Eq:Q}) into Eq. (\ref{Eq:J-tot}) we obtain,
\begin{equation}
\label{Eq:J-final}
{\cal J}_z=-\frac{n^2\hbar\theta}{8\pi}, 
\end{equation}
which for $\theta=\pi$, corresponding to TIs with either time-reversal or inversion symmetries, yields ${\cal J}_z=-n^2\hbar/8$. The 
latter corresponds to twice the value obtained for a model of self-dual CS vortices \cite{Jackiw-Weinberg} (see also SI). Note that in that case there 
is no Maxwell term and therefore only the mechanical angular momentum contributes. 

Equation (\ref{Eq:J-final}) implies that the Cooper pair-vortex composites in the TI-SC interface obey anyonic statistics, since the angular momentum is neither integer nor half-integer. The statistics implied by the angular momentum (and a consistency check on possible ${\cal J}_z$ values) may be rationalized by the AB phases associated with the dyon electric charge ($Q$) and magnetic field flux ($\Phi$) components. Let $|\Psi({\bf r}),\Psi(-{\bf r})\rangle$ represent a product state of two identical dyons (each of the same angular momentum) % (an axial vector) $\mathbfcal{J}$)
in their center of mass frame. The total $z-$component of the angular momentum of the dyons $(2{\cal{J}}_{z}$) may generate a rotation $R_{\pi}$ of the pair by $\pi$ about the origin (effectively exchanging the two dyons with one another). This leads to a statistical transmutation (braiding) phase of $R_{\pi} |\Psi({\bf r}),\Psi(-{\bf r})\rangle = e^{i \pi \Theta} |\Psi({\bf r}),\Psi(-{\bf r})\rangle$ with $\Theta \equiv 2{\cal J}_z/\hbar$. The latter phase factor is the Dirac phase associated with $R_{\pi}$ or, equivalently, the AB phase for a full rotation of one dyon around the other, i.e., $e^{i Q \Phi/(\hbar c)}$. For TI-SC interfaces with dyons of charge $Q= - ne\theta/(4 \pi)$ and flux $\Phi= n \Phi_{0}$, the equivalence $e^{i \pi \Theta}  = e^{i Q \Phi/(\hbar c)}$ leads to quantized ${\cal{J}}_{z}$ values consistent with our central result of Eq. (\ref{Eq:J-final}). We stress that the above considerations are independent of the system geometry, thus demonstrating the generality of our results.

\begin{figure}
	\centering
	\begin{tabular}{l l}
		\hspace{0cm}(a) &
		\hspace{0cm}(b) \\
		\includegraphics[width=0.52\columnwidth]{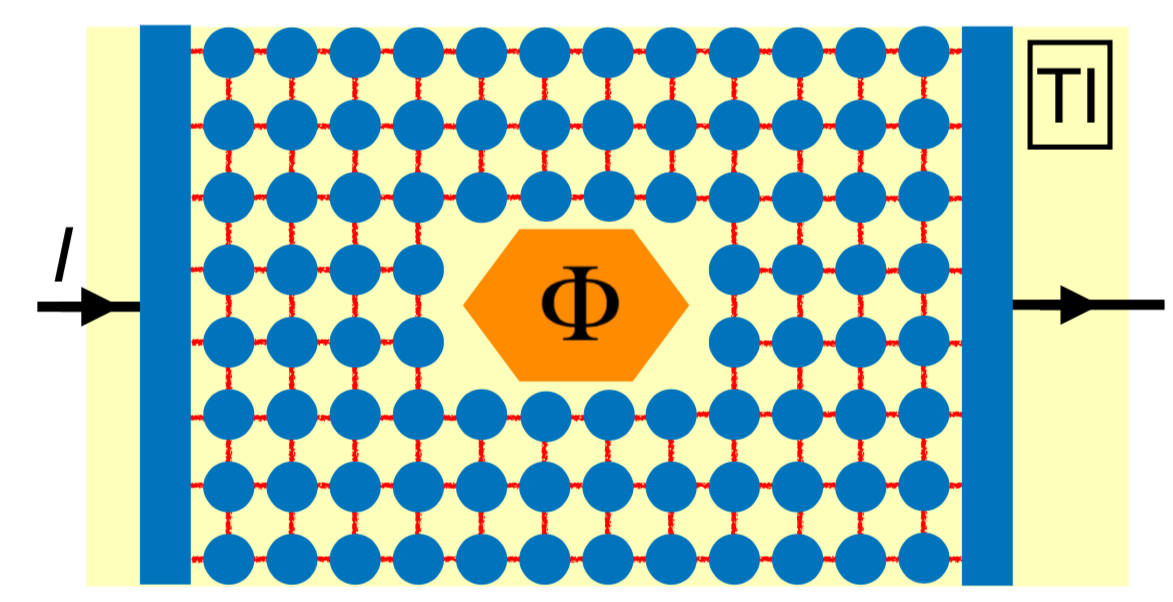} &
		\includegraphics[width=0.47\columnwidth]{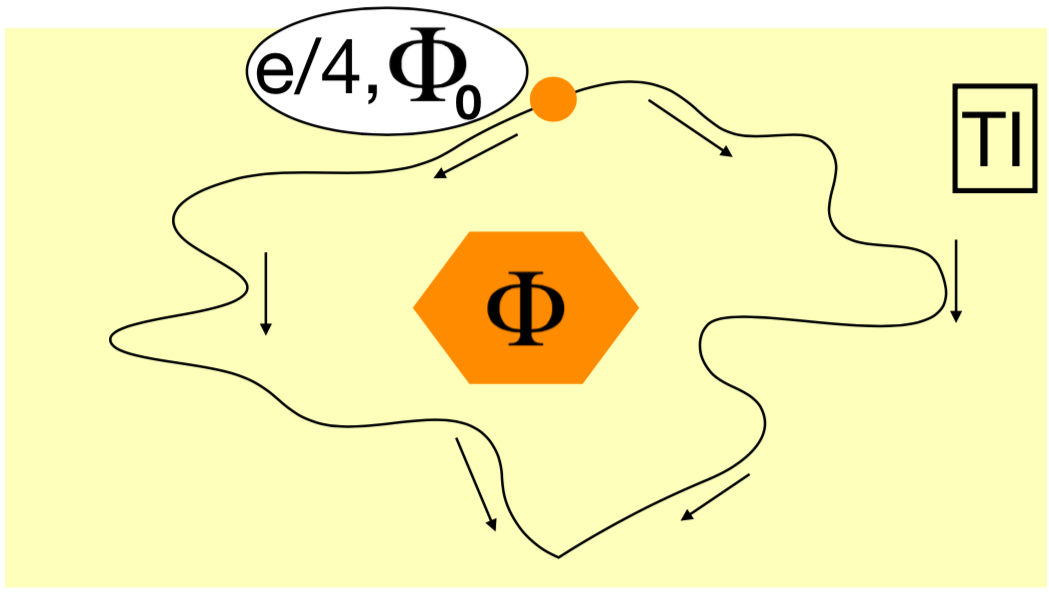} \\
		\end{tabular}
	\caption{(a) Sketch of the Josephson junction array on a TI to measure the anyon fractional statistics. (a) Circles represent superconducting islands and lines for Josephson weak links connecting these islands. The bias current I is injected to the left electrode and collected from the right electrode of the array. (b) Illustration of anyon with charge $e/4$ and elementary flux $\Phi_0$ moving between plaquettes that interferes when passing a flux island with total flux $\Phi$. 
	}
	\label{Fig:JJarray}
\end{figure}

{\it Experimental detection of fractional statistics ---} 
As we explained above, our computed total angular momentum can be rationalized via a calculation of the AB phase. Similarly, the AB phase can be evaluated from the total (electromagnetic field + mechanical) angular momentum ${\cal J}_z$. Thus, our found ${\cal J}_z$ values imply corresponding results for the AB phase. Similarly, experimentally measured AB type phases may also determine, up to modular corrections, and test our predictions for the total angular momentum. Our analysis of the AB phases associated with the dyon electric charge and magnetic field flux components indeed suggests that the ensuing fractional statistics is observable via an interference experiment. Interestingly, interference of fluxes has been demonstrated via the Aharonov-Casher effect for vortices in Josephson-junction (JJ) arrays at the surface of a trivial insulator already 25 years ago \cite{Elion1993}, following the theoretical works in Refs. \cite{Resnik-Aharonov_PhysRevD.40.4178,Van-Wees_PhysRevLett.65.255}.  
This experiment measured the phase difference accumulated by magnetic fluxes as they pass on either side of a charged island.   %(Elion, Wachters, Sohn and Mooij, PRL 71, 14, 2311 (1993)).
When manufactured on a TI surface, fluxes in the plaquettes of the JJs pierce the TI and thus will attain dyon statistics. In the flux flow regime, a similar phase accumulation occurs for anyons passing either side of a flux island (Fig. \ref{Fig:JJarray}). Due to interference and the quantized ${\cal{J}}_{z}$ values of the dyons, we predict an 8$\pi$ periodicity in the differential resistance of the JJ array. This periodicity is expected to be robust. It realizes a fingerprint of (Abelian) anyon statistics in absence of any superconducting proximity effect in the TI.

%Although we have not explicitly considered the fermions in our analysis,  the above results are further consistent with the existence of half-vortices  in $p$-wave superconductors \cite{Ivanov,Volovik}, which are predicted to be induced by proximity of an $s$-wave SC with a TI \cite{Fu-Kane-2008}.  Indeed, this case is known to host zero mode bound state Majorana fermions at the vortices. Two such vortices correspond Majorana modes $\gamma_1$ and $\gamma_2$, which can be combined to form the fermion operator, $c=(\gamma_1+i\gamma_2)/2$.  Thus, the operator $e^{\pi\gamma_2\gamma_1/4}$ exchanging two Majorana vortices is given by $e^{i\pi(2c^\dagger c-1)/4}=e^{\pm i\pi/4}$ \cite{Ivanov}, which is consistent with our expression for the statistical angle for $n=1$. 

%{\it Summary ---} 

In summary, we demonstrated that the axion term of the topological insulator endows impinging magnetic vortices from a superconductor with a nontrivial fractionalized angular momentum. The magnetic vortex-electric charge hybrids forming at the topological insulator interfaces may thus constitute a long sought tabletop realization of anyons in a new and rather accessible experimental arena.

\begin{acknowledgments}
{\it Acknowledgments ---}
We thank Marcel Franz and Alexander Brinkman for fruitful discussions.
	FSN acknowledges the support of the Priority Program SPP 1666 of the German Research Foundation (DFG) under grant no. ER 463/9. 
	This work was supported by the DFG through the Collaborative Research Center SFB 1143. 
	ZN acknowledges support by the NSF under grant no. CMMT 1411229. 
\end{acknowledgments}

\bibliography{axion-anyon-vortex-1}

\section{Supplemental Information}
\renewcommand\theequation{S\arabic{equation}}
\setcounter{equation}{0}

\subsection{Chern-Simons vortices}

For comparison we briefly summarize the theory for self-dual CS vortices by Jackiw and Weinberg \cite{Jackiw-Weinberg}. 
This model can be understood in relatively simple terms due to the absence of a Maxwell term in the Lagrangian.

The topological current $j_\mu=\sigma_{H}\epsilon_{\mu\nu\lambda}\partial^\nu A^\lambda$ 
also yields the electromagnetic response of the system. Here  $\sigma_H=e^2/(2h)$ is the half-quantized 
Hall conductivity and $A^\mu$ is the electromagnetic gauge potential with the Greek indices being associated to 
spacetime coordinates in 2+1 dimensions.  Since  
the zeroth component of $j^\mu$ represents the charge density we have $j_0=c\rho=\sigma_H\epsilon_{ij}\partial_iA_j=\sigma_HB$. 
Thus, if ${\bf A}$ is the vector potential associated to a 
superconducting vortex, we have the magnetic flux, $\Phi=n\Phi_0$, where $n$ is an integer and $\Phi_0=hc/(2e)$ is the elementary flux 
quantum. It follows that, 
\begin{equation}
\label{Eq:vortex-charge}
Q=\int d^2r\rho({\bf r})=\frac{\sigma_H}{c}\int d^2rB({\bf r})=\frac{n\Phi_0\sigma_H}{c}=\frac{ne}{4}.
\end{equation}

Since in this case there is no Maxwell term, the angular momentum is given just by the mechanical one. The calculation proceeds in a 
similar way as in our main text, except that now we are dealing with two-dimensional system. Thus, we have, 
\begin{equation}
J=\frac{e}{2\hbar c}\int_0^\infty dr r\left[\frac{2e}{c}rA(r)-n\hbar\right]B(r).
\end{equation} 
Since, 
\begin{equation}
A(r)=\frac{n\Phi_0}{2\pi r}\left[1-m_LrK_1(m_Lr)\right], 
\end{equation}
we obtain, 
\begin{equation}
J=-\frac{m_L^3n^2\hbar}{8}\int_0^\infty dr r^2K_0(m_Lr)K_1(m_Lr)=-\frac{n^2\hbar}{16}.
\end{equation}

\subsection{Expression for $f(p)$}
The boundary condition (11) allows to determine $f(p)$ as
\begin{equation}
\label{Eq:f}
f(p)=\frac{p\left(p+\epsilon \sqrt{p^2+m_L^2}\right)+m_L^2}{p\left\{[1+\epsilon+(\alpha\theta/\pi)^2]p+(1+\epsilon)\sqrt{p^2+m_L^2}\right\}
	+m_L^2}.
\end{equation}

Note that $f(0)=1$ and that $f(p)$ is independent of $\epsilon$ for $\theta=0$,
\begin{equation}
\label{Eq:f-1}
f(p)=\frac{\sqrt{p^2+m_L^2}}{p+\sqrt{p^2+m_L^2}}.
\end{equation}
Since $(\alpha\theta/\pi)^2\ll 1$, the above expression is actually very accurate. 

\subsection{Explicit expressions of the electromagnetic fields}

From Eqs. (12) and (15) we obtain the analytic expressions for the components of the electric and magnetic fields in 
cylindrical coordinates, 
\begin{equation}
\label{Eq:Br}
B_r(r,z)=-\frac{n\Phi_0m_L^2}{2\pi}\int_0^\infty dp\frac{J_1(pr)}{p^2+m_L^2}\frac{\partial a(p,z)}{\partial z},
\end{equation}
\begin{equation}
\label{Eq:Bz}
B_z(r,z)=\frac{n\Phi_0m_L^2}{2\pi}\int_0^\infty dp\frac{pJ_0(pr)a(p,z)}{p^2+m_L^2}, 
\end{equation}
\begin{equation}
\label{Eq:Er}
E_r(r,z)=\frac{ne\theta m_L^2}{2\pi}\int_0^\infty dp\frac{p^2J_1(pr)F(p,z)}{\left(\sqrt{p^2+m_L^2}+\epsilon p\right)(p^2+m_L^2)},
\end{equation}
\begin{equation}
\label{Eq:Ez}
E_z(r,z)=-\frac{ne\theta m_L^2}{2\pi}\int_0^\infty dp\frac{pJ_0(pr)\partial F(p,z)/\partial z}{\left(\sqrt{p^2+m_L^2}+\epsilon p\right)(p^2+m_L^2)},
\end{equation}
where $J_k(u)$ is a Bessel function of first kind. We verify that $\nablab\times{\bf B}=0$ for $z>0$, as it should \cite{Brandt-1981}, 
implying that ${\bf B}=\nablab\chi$ in this case, where, 
\begin{equation}
\chi(r,z)=-\frac{n\Phi_0m_L^2}{2\pi}\int_0^\infty dp\frac{J_0(pr)e^{-pz}f(p)}{p^2+m_L^2}.
\end{equation}

\subsection{Finiteness of the electric energy}

The magnetic energy per unit length of the vortex line is obviously finite, since the expression for the magnetic field does not differ 
appreciably from the usual one for an ANO vortex. On the other hand, for the electric energy we have mentioned in the main text that 
the main obstacle preventing charged vortices within a non-topological  ANO model to exist relies on the lack of a finite electric energy per unit length, which is the result obtained originally by Julia and Zee \cite{Julia-Zee}. Here we show that this energy density is finite. 

The energy density is given for $z<0$ by, 
\begin{equation}
{\cal E}_{\rm el}=\frac{1}{8\pi}\int d^2r{\bf E}^2(r,z).
\end{equation}
Inserting Eqs. (\ref{Eq:Er}) and (\ref{Eq:Ez}) in the above equation and using the identity, 
\begin{equation}
\int_{0}^{\infty}dr rJ_n(pr)J_n(qr)=\frac{1}{p}\delta(p-q),
\end{equation}
and subsequently integrating out $q$, we obtain, 
\begin{eqnarray}
{\cal E}_{\rm el}&=&\frac{n^2e^2\theta^2m_L^4}{32\pi^3}\int_{0}^{\infty}dp\frac{pf^2(p)e^{2z\sqrt{p^2+m_L^2}}}{\left(\sqrt{p^2+m_L^2}+\epsilon p\right)^2(p^2+m_L^2)}\nonumber\\
&\times&\left(1+\frac{p^2}{p^2+m_L^2}\right).
\end{eqnarray}
The above integral is clearly finite, being maximal at $z=0$.

\end{document}